\documentclass[letterpaper, 10 pt, conference]{ieeeconf}

\IEEEoverridecommandlockouts 

\usepackage[caption=false]{subfig}
\usepackage{cite}
\usepackage{graphicx}
\usepackage{multirow}
\usepackage{algorithmic}
\usepackage[utf8]{inputenc}
\usepackage{amsmath}
\usepackage{CJKutf8}
\usepackage{tabularx}
\usepackage{booktabs}
\usepackage{makecell}

\title{\LARGE \bf
Efficient 4D Radar Data Auto-labeling Method using \\LiDAR-based Object Detection Network}

\author{Min-Hyeok Sun$^{\dag}$, Dong-Hee Paek$^{\dag}$, Seung-Hyun Song and Seung-Hyun Kong$^{*}$% <-this % stops a space $^{\dag}$
\thanks{$^{\dag}$ co-first authors, * corresponding author}% <-this % stops a space
\thanks{The authors are with Graduate School of Mobility, Korea Advanced Institute of Science and Technology(KAIST), 193, Munji-ro, Yuseonggu, Daejeon, Republic of Korea{\tt\small \{hyeok0809, donghee.paek, shyun, skong@kaist\}.ac.kr}}}

\begin{document}

\maketitle
\thispagestyle{empty}
\pagestyle{empty}

%%%%%%%%%%%%%%%%%%%%%%%%%%%%%%%%%%%%%%%%%%%%%%%%%%%%%%%%%%%%%%%%%%%%%%%%%%%%%%%%
\begin{abstract}
Focusing on the strength of 4D (4-Dimensional) radar, research about robust 3D object detection networks in adverse weather conditions has gained attention. To train such networks, datasets that contain large amounts of 4D radar data and ground truth labels are essential. However, the existing 4D radar datasets (e.g., K-Radar) lack sufficient sensor data and labels, which hinders the advancement in this research domain. Furthermore, enlarging the 4D radar datasets requires a time-consuming and expensive manual labeling process. To address these issues, we propose the auto-labeling method of 4D radar tensor (4DRT) in the K-Radar dataset. The proposed method initially trains a LiDAR-based object detection network (LODN) using calibrated LiDAR point cloud (LPC). The trained LODN then automatically generates ground truth labels (i.e., auto-labels, ALs) of the K-Radar train dataset without human intervention. The generated ALs are used to train the 4D radar-based object detection network (4DRODN), Radar Tensor Network with Height (RTNH). The experimental results demonstrate that RTNH trained with ALs has achieved a similar detection performance to the original RTNH which is trained with manually annotated ground truth labels, thereby verifying the effectiveness of the proposed auto-labeling method. All relevant codes will be soon available at the following GitHub project: https://github.com/kaist-avelab/K-Radar
\end{abstract}
\vspace{10mm}
%%%%%%%%%%%%%%%%%%%%%%%%%%%%%%%%%%%%%%%%%%%%%%%%%%%%%%%%%%%%%%%%%%%%%%%%%%%%%%%%
%%%%%%%%%%%%%%%%%%%%%%%%%%%%%%%%%%%%%%%%%%%%%%%%%%%%%%%%%%%%%%%%%%%%%%%%%%%%%%%%

\section{INTRODUCTION}

As one of the fundamental modules of autonomous driving systems, perception modules enable vehicles to identify various types of driving information \cite{levinson2011towards}. Among perception modules, object detection identifies the type, shape, and position of objects around the vehicle. Robust 3D (3-Dimensional) object detection under all weather conditions is required to ensure safe driving since autonomous driving systems subsequently perform path planning and control based on the outputs of 3D object detection \cite{gupta2024robust}.

Camera and LiDAR are primary sensors for object detection. Camera can provide high-resolution RGB images for accurate 2D object detection \cite{feng2021review}. However, camera lacks depth information, which is essential for precise 3D object detection. Furthermore, since camera measures visible light, it is difficult to robustly detect objects under adverse illumination or weather conditions. LiDAR measures the time-of-flight of infrared pulse signals and outputs precise 3D spatial information of the surrounding environment through LiDAR point cloud (LPC), which enables accurate 3D object detection \cite{wu2020deep}. In addition, since LiDAR utilizes infrared pulse signals, they are less affected by illumination conditions than camera. However, LiDAR sensors are susceptible to raindrops or snowflakes, which degrades the detection performance of LiDAR-based object detection networks (LODNs) in adverse weather conditions \cite{vargas2021overview}.

\begin{figure}[t]
\vspace{0.2cm}
\centerline{\includegraphics[width=\linewidth]{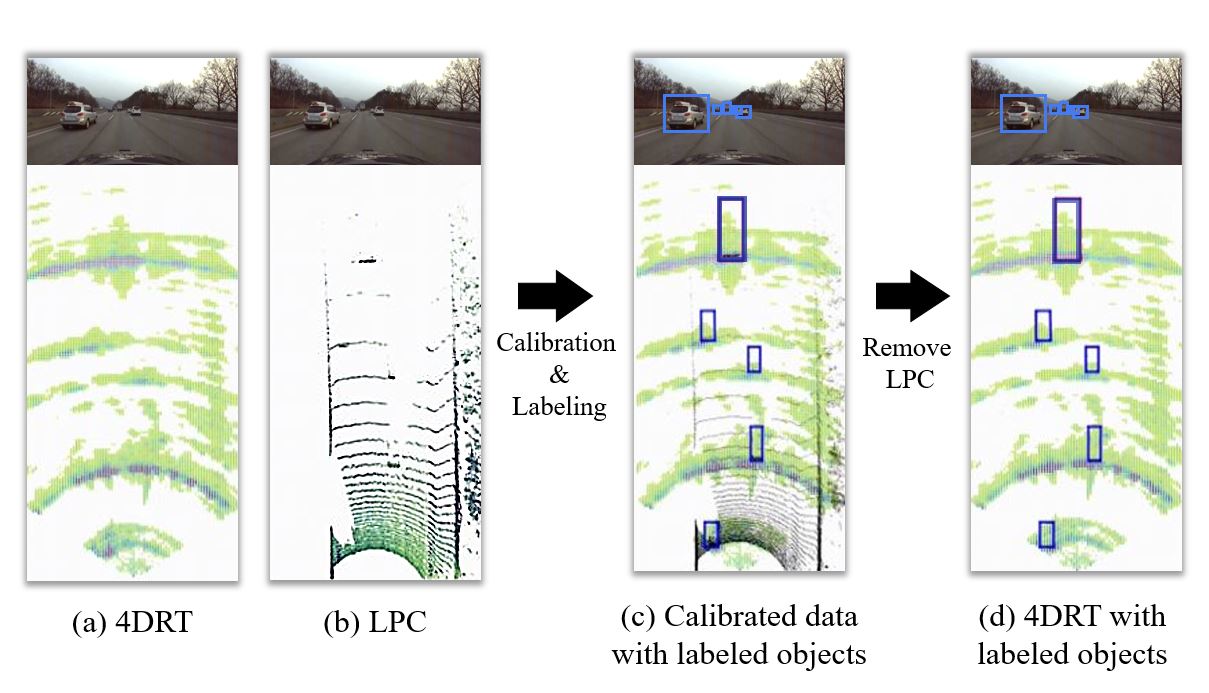}}
\caption{Overview of 4DRT labeling process. 4DRT is not intuitive for human eyes to distinguish various objects in the data. In contrast, LPC can represent the detailed boundaries of objects, which can be utilized as clues to annotated ground truth labels. Therefore, calibration of 4DRT with LPC enables labeling of 4DRT by leveraging the detailed object representations from LPC.}
\label{fig_calibration}
\end{figure}

Recently, 4D radar has emerged as an alternative sensor for robust object detection under adverse illumination and weather conditions \cite{sun20214d}. As radar utilizes long-wavelength radio waves, 4D radar is less affected by illumination and weather conditions than camera or LiDAR \cite{bilik2022comparative}. Furthermore, unlike 3D radar which only can measure along the azimuth, range, and Doppler dimensions, 4D radar additionally can measure along elevation dimension. Therefore, 4D radar can recognize the 3D shapes of objects similar to LiDAR. Focused on the application of 4D radar to autonomous driving systems, studies \cite{astyx, rpfanet, vod, tj4dradset, k-radar, e-k-radar} published 4D radar dataset or proposed 4D radar-based object detection networks (4DRODNs). Among them, \cite{k-radar} and \cite{e-k-radar} released a large-scale multi-modal 4D radar dataset (i.e., K-Radar), which contains 4D radar tensor (4DRT) captured under various illumination, weather, and road conditions. The studies also proposed Radar Tensor Network with Height (RTNH), one of the most practical 4DRODNs trained with the K-Radar dataset. RTNH demonstrated the robust detection performance in adverse weather conditions such as sleet and heavy snow than LODN, which verified the strength of the application of 4D radar in autonomous driving systems.

\newpage
However, the overall detection performance of 4DRODNs is still not enough to ensure safe autonomous driving \cite{jin2023radar}. One fundamental reason is the insufficient amount of available datasets for training. Although 4D radar has been commercialized, there are only a few datasets available recently. In addition, the amount of sensor data and manually annotated ground truth labels (i.e., handmade labels, HLs) in each dataset is small compared to published LiDAR datasets. For example, K-Radar, the largest 4D radar dataset so far, contains about 100K labeled objects, which is minor compared to the LiDAR datasets\cite{kitti, nuscenes, waymo} (e.g., 200K for KITTI\cite{kitti}, 1.4M for nuScenes\cite{nuscenes}, 12M for the Waymo open dataset\cite{waymo}). Therefore, the quantitative expansion of the K-Radar dataset is essential for advanced 4DRODN research. Unfortunately, expanding or generating datasets is a very expensive task in terms of both time and cost \cite{dvornik2018modeling}. In addition, as shown in Fig.\ref{fig_calibration}(a), 4DRT is not as intuitive to human eyes as other sensor data (RGB image or LPC), which impedes manually labeling the precise 3D bounding boxes of objects in 4DRT.

In this paper, we propose an auto-labeling method to address these issues. The proposed auto-labeling method exploits a LODN, which is trained with LPC data calibrated with 4DRT. The method automatically generates ground truth labels (i.e., auto-labels, ALs) of the 17.5K K-Radar train dataset without additional human labor or expense. In the experiment, we train RTNH with ALs (i.e., RTNH-AL) and compare the detection performance with the original RTNH trained with HLs. The result verifies the efficiency and effectiveness of our proposed auto-labeling method. We further conduct additional experiments and ablation studies to seek the strategies for efficient K-Radar expansion using the proposed auto-labeling method. In summary, the contributions of this paper are as follows:

\begin{itemize}
\item We propose an auto-labeling method that automatically generates ground truth labels. The proposed method facilitates the efficient expansion of the 4D radar dataset (e.g., K-Radar) without additional human intervention or labor costs.

\item We train the RTNH using generated ALs and demonstrate a comparable detection performance with the RTNH trained with HLs. The result verifies the effectiveness of the proposed auto-labeling method.

\item We also present the strategies for efficient K-Radar expansion using the proposed auto-labeling methods based on the experiment results. \\
\end{itemize}

%%%%%%%%%%%%%%%%%%%%%%%%%%%%%%%%%%%%%%%%%%%%%%%%%%%%%%%%%%%%%%%%%%%%%%%%%%%%%%%%
%%%%%%%%%%%%%%%%%%%%%%%%%%%%%%%%%%%%%%%%%%%%%%%%%%%%%%%%%%%%%%%%%%%%%%%%%%%%%%%%

\vspace{2mm}
\section{Related works}
Recent advancements in radar hardware technology has led the active commercialization of 4D radar sensors \cite{sichani2023antenna}.  The active commercialization has enabled the use of 4D radar in autonomous driving perception, and several 4DRODN studies and 4D radar datasets have been published.

ASTYX\cite{astyx} is the first published 4D radar dataset that includes 0.5K frames of the 4D radar point cloud (4DRPC) and ground truth labels. Study \cite{rpfanet} proposed RPFA-net, which is similar to LODN PointPillars \cite{pointpillars}. The authors trained RPFA-net with ASTYX dataset and demonstrated the first 4D radar-based 3D object detection. Although these studies are notable for pioneering the utilization of 4D radar in autonomous driving systems, the dataset was too small to develop and train advanced networks. Studies \cite{vod} and \cite{tj4dradset} published VoD and TJ4DRadSet datasets that contain 8.7K and 7.8K frames of 4DRPC, respectively, and performed 4D radar based 3D object detection using PointPillars network. However, they only captured 4DRPC under good weather conditions. Therefore, these studies can not explore the strength of 4DRODNs, which can robustly detect objects in adverse weather conditions. In addition, these datasets are still insufficient to train practical 4DRODNs for autonomous driving systems.

In contrast, study \cite{k-radar} released a K-Radar dataset that provides 35K frames of 4D radar data captured in diverse road, illumination, and weather conditions. In addition, the K-Radar dataset offers 4D radar tensor (4DRT) instead of 4DRPC. 4DRT is a raw-level radar measurement before Constant False Alarm Rate (CFAR) process. Therefore, 4DRT can provide denser 3D spatial information than the 4DRPC. The study also proposed a practical and real-time operating 4DRODN, RTNH. The experiment results in \cite{k-radar} demonstrated that RTNH has superior detection performance in challenging weather conditions compared to LODNs. A follow-up study \cite{e-k-radar} introduced a pre-processing method to filter 4DRT and achieved enhanced detection performance.

However, the overall detection performance of RTNH is not enough to ensure safe autonomous driving. One primary reason is that the K-Radar train dataset does not contain sufficient labeled objects compared to the LiDAR datasets. For example, K-Radar has 100K annotated objects, whereas KITTI\cite{kitti}, nuScenes\cite{nuscenes}, and Waymo\cite{waymo} provide 200K, 1.4M, and 12M labeled objects, respectively. In general, a large amount of labeled objects is crucial to train advanced object detection networks \cite{he2022partimagenet}. Therefore, efficient ways to expand the K-Radar dataset are necessary.\\

\begin{figure}[!t]
\vspace{0.2cm}
\centerline{\includegraphics[width=\linewidth]{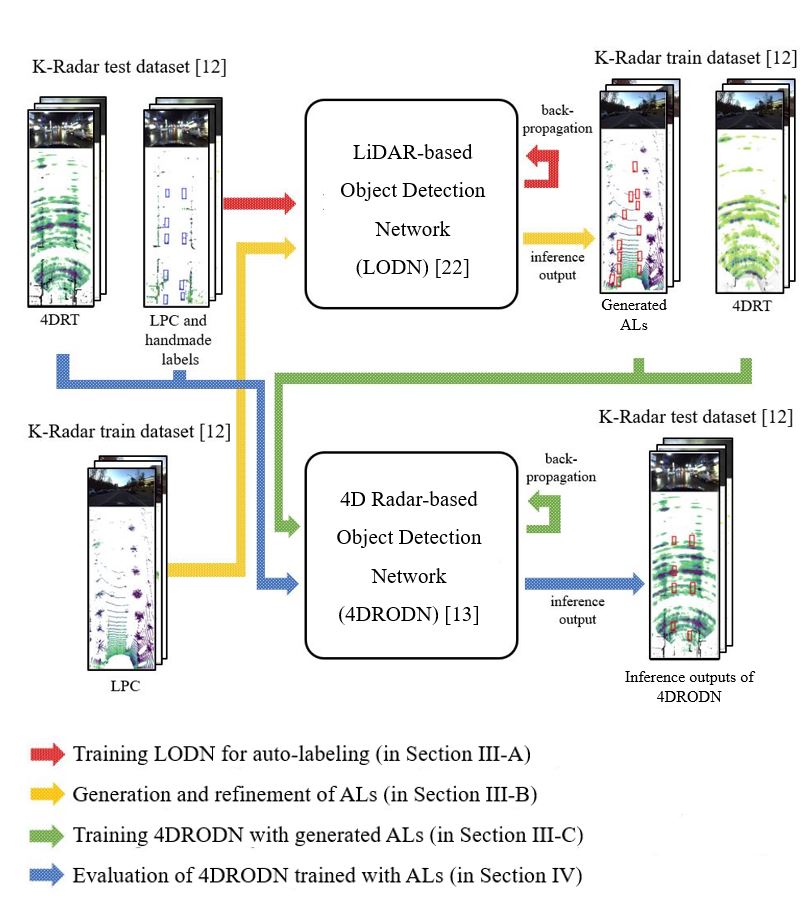}}
\caption{The overall process of the proposed auto-labeling method. The proposed method trains LODN (PVRCNN++ \cite{pvrcnn++}) using calibrated LPC and HLs of the K-Radar test dataset. The method then generates ALs of the K-radar train dataset by saving inference outputs of trained LOND. The generated ALs are used to train 4DRODN (RTNH\cite{e-k-radar}) to verify the effectiveness of the proposed method.}
\label{overall_structure}
\vspace{-0mm}
\end{figure}

%%%%%%%%%%%%%%%%%%%%%%%%%%%%%%%%%%%%%%%%%%%%%%%%%%%%%%%%%%%%%%%%%%%%%%%%%%%%%%%%
%%%%%%%%%%%%%%%%%%%%%%%%%%%%%%%%%%%%%%%%%%%%%%%%%%%%%%%%%%%%%%%%%%%%%%%%%%%%%%%%

\section{Proposed Auto-labeling method}

4DRT contains raw power measurements of the radio signal reflected from objects, but there are also invalid measurements from the sidelobe, noise, interference, and clutter. Therefore, as shown in Fig.\ref{fig_calibration}, 4DRT is not intuitive for human eyes to distinguish various types of objects, such as vehicles, pedestrians, road signs, or curbs\cite{k-radar}. On the other hand, LPC can represent the detailed boundaries of objects, and is easy to distinguish object types in the data. Therefore to generate HLs of K-Radar dataset, \cite{k-radar} accurately calibrated the 4DRT to align with the LPC. Subsequently, they used the LPC object representations as a reference and generated ground truth labels for 4DRT manually. Similar to the manual labeling process, after training LODN to recognize objects on the LPC, the inference outputs of the trained LODN can serve as ALs for the 4DRT.

The overall process of the proposed auto-labeling method is shown in Fig.\ref{overall_structure}. The proposed method initially trains LODN to perform object detection using LPC data. The trained LODN is then applied to the LPC of the K-Radar train dataset to generate ALs. The method additionally conducts an additional refinement process to obtain the accurate ALs. Finally, the generated ALs are used to train the 4DRODN (RTNH). The following subsections explain the detail of each process.

\subsection{Training LODN for Auto-labeling}

The proposed auto-labeling method introduces PVRCNN++\cite{pvrcnn++} to train LODN. The PVRCNN++ is a point-voxel-based LODN that initially proposes 3D bounding box candidates after applying the 3D sparse convolution networks to voxelized LPC as in \cite{second}. PVRCNN++ then applies PointNet++\cite{pointnet++} to extract key point features of the bounding box candidates, which are then utilized to refine the bounding box candidates and output precise final detection results.

Since the K-Radar train dataset will be used later to train 4DRODN, the PVRCNN++ should not be overfitted to the K-Radar train dataset. Otherwise, it is hard to determine the generalized effectiveness of the proposed auto-labeling method. Therefore, the proposed method trains PVRCNN++ with the K-Radar test dataset instead of the train dataset to prevent overfitting when auto-labeling the remaining K-Radar train dataset.

\subsection{Generation and Refinement of ALs} \label{section_refinement}

After training PVRCNN++, The proposed method generates ALs of the K-Radar train dataset with the inference output of PVRCNN++. When generating ALs, the confidence score threshold significantly affects the accuracy of the generated ALs. The confidence score represents the confidence level for each detected object in the inference output of the trained object detection network. Therefore, a lower confidence score threshold reduces miss detections but increases false alarms in the generated ALs and vice versa. To determine the optimal threshold, the method generates multiple ALs with three different thresholds and compares the accuracy of ALs based on the \textit{F1}-score metric. The comparison result in Table \ref{confidence threshold} demonstrates that a threshold of 0.3 is the optimal.

\begin{table}[!h]
\caption{Comparison of generated ALs accuracy based on the confidence score threshold}
\begin{tabular}{c*{4}{>{\centering\arraybackslash}p{1.65cm}}}
\hline
\begin{tabular}[c]{@{}c@{}}\textbf{Confidence score}\\ \textbf{threshold}\end{tabular} & \textbf{Precision} & \textbf{Recall} & \textbf{\textit{F1}-score} \\
\hline
\hline
0.1 & 0.757 & 0.621 & 0.683 \\
0.3 & 0.878 & 0.593 & 0.708 \\
0.5 & 0.930 & 0.557 & 0.697 \\
\hline
\end{tabular}
\label{confidence threshold}
\vspace{1mm}
\end{table}

Even after a selection of the optimal confidence score threshold, the generated ALs still contain a number of inaccurate objects. In some cases, these inaccuracies arise from the limited detection capabilities of the trained LODN. However, there are cases where only the current frame (frame at $t$) shows incorrect detection but not in temporal neighboring frames (frames at $t-1$ and $t+1$). To refine these intermittent inaccurate objects, we introduced a refinement process. Specifically, the process compares the intersection of unions (IoU) of all the auto-labeled objects in frame at t with those of the objects in frames at $t-1$ and $t+1$. Subsequently, the process finds the miss detections and false alarms based on the IoU threshold. Miss detections in the frame at $t$ are supplemented with the mean values of two 3D bounding boxes in labels of frames at $t-1$ and $t+1$. Conversely, false alarms in the frame at $t$ are filtered out during the refinement process. We conduct an ablation study to confirm the effect of the refinement process on RTNH-AL detection performance in subsection \ref{exp_a_2}.

\subsection{Training 4DRODN with Generated ALs} \label{3.c}

The proposed method introduces RTNH\cite{e-k-radar} to train 4DRODN with generated ALs. In \cite{e-k-radar}, RTNH is trained with 4D sparse radar tensor (4DSRT), which includes only the top 10\% measurements of the 4DRT, and the experiment results demonstrate the enhanced detection performance. Similarly, we train RTNH (i.e., RTNH-AL) using 4DSRT and ALs, to compare with the original RTNH and verify the effectiveness of the proposed auto-labeling method in subsection \ref{section_exp_1}. 

Additionally, we hypothesize that the performance degradation of LODN in adverse weather conditions would negatively impact the training of RTNH-AL. LiDAR cannot output reliable LPC in adverse weather conditions, which degrades the detection performance of the LODN. Such degradation is evident in Table \ref{exp_3}, where LODNs (SECOND\cite{second} and PVRCNN++\cite{pvrcnn++}) demonstrate a reduced detection performance in sleet and heavy snow conditions compared to other weather conditions. The performance degradation may promote the generation of inaccurate ALs, which can negatively affect the RTNH-AL detection performance.

In contrast, 4D radar can output reliable 4DRT robustly across various weather conditions. This implies that, unlike LPC, the data distribution of 4DRT under adverse weather conditions is not different from that under normal weather conditions. Consequently, we expect that such a consistent data distribution of 4DRT enables the generalized training of RTNH-AL across various weather conditions. In other words, we can expect the generalized detection performance of RTNH-AL while the network is trained without the 4DRT and ALs that may contain multiple inaccurate objects in adverse weather conditions. 

To explore such generalized training of RTNH-AL, we divide the auto-labeled K-Radar train dataset into three subsets and train the RTNH-AL models using each subset in subsection \ref{section_exp_2}. Specifically, we train the RTNH-AL-NO model with the subset that contains only ideal weather conditions, such as normal and overcast (i.e., NO). Next, we train the RTNH-AL-NOFRL model with the subset that contains normal, overcast, fog, rain, and light snow (i.e., NOFRL) weather conditions where there are light raindrops and snowflakes but no performance degradation is evident in PVRCNN++. Lastly, we train the RTNH-AL-ALL model with the subset that contains all weather conditions, including adverse weather conditions such as sleet and heavy snow.\\

%%%%%%%%%%%%%%%%%%%%%%%%%%%%%%%%%%%%%%%%%%%%%%%%%%%%%%%%%%%%%%%%%%%%%%%%%%%%%%%%
%%%%%%%%%%%%%%%%%%%%%%%%%%%%%%%%%%%%%%%%%%%%%%%%%%%%%%%%%%%%%%%%%%%%%%%%%%%%%%%%

\section{Experiment results}

In this section, we present the experiment results of training RTNH using the proposed auto-labeling method. In subsection \ref{section_exp_1}, we compare the detection performance of RTNH-AL (trained with ALs) with that of the original RTNH (trained with HLs) and demonstrate the effectiveness of the proposed auto-labeling method. Subsequently, in subsection \ref{section_exp_2}, we present the results of three RTNH-AL models trained with three divided subsets and discuss the generalized training of RTNH-AL. Finally, we conduct ablation studies in subsection \ref{section_exp_3} to explore ways to improve the training effectiveness of RTNH using the auto-labeling method further based on quantitative experimental results.

All the training environments and hyper-parameters are set identically to those in \cite{e-k-radar} (implemented with PyTorch 1.11.0 on Ubuntu 20.04 machines equipped with RTX 3090). We evaluate the both bird's-eye view (BEV) and 3D detection performance of trained RTNH models with the metric of IoU-based average precision ($\textit{AP}_{\textit{BEV}}$ and $\textit{AP}_{\textit{3D}}$ respectively) for sedan objects with a 0.3 IoU threshold, which follows the metric in \cite{k-radar}.

\subsection{Comparison of RTNH-AL and RTNH} \label{section_exp_1}

\begin{figure}[!t]
\vspace{0.2cm}
\centerline{\includegraphics[width=\linewidth]{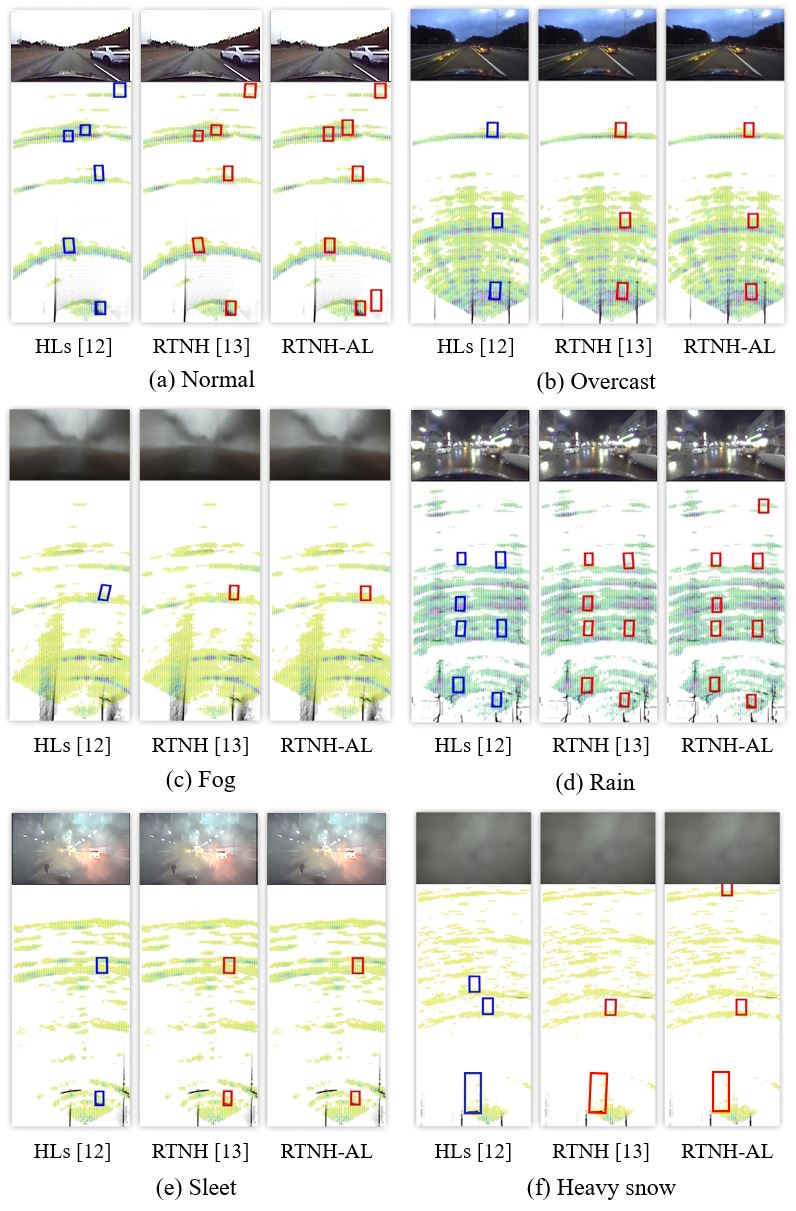}}
\caption{Comparison of RTNH and RTNH-AL inference outputs in various weather conditions. In each example frame, HLs, inference outputs of RTNH, and those of RTNH-AL are presented on the 4DRT. The blue boxes represent the ground truth objects in HLs and the red boxes represent the detected objects of RTNH models..}
\label{fig_RTNH_AL}
\end{figure}

Table \ref{exp_1} presents the $\textit{AP}_{\textit{BEV}}$ and $\textit{AP}_{\textit{3D}}$ for car objects of RTNH and RTNH-AL. As shown in the table, the RTNH-AL demonstrates almost the same level of detection performance as that of RTNH. This result verifies that the proposed auto-labeling method that utilized LODN trained with LPC is desirable to train RTNH efficiently. The example of inference output is shown in Fig.\ref{fig_RTNH_AL} where the HLs, inference output of RTNH, and RTNH-AL of example frames under various weather conditions are presented.  

\begin{table}[!h]
\caption{Comparison of RTNH and RTNH-AL detection performance}
\centering
\begin{tabular}{c*{10}{>{\centering\arraybackslash}p{0.36cm}}}
\hline
\fontsize{6}{7}\selectfont{\textbf{Model}} & \fontsize{6}{7}\selectfont{\textbf{Met- ric}} & \fontsize{6}{7}\selectfont{\textbf{Over- all}} & \fontsize{6}{7}\selectfont{\textbf{Nor- mal}} & \fontsize{6}{7}\selectfont{\textbf{Over- cast}} & \fontsize{6}{7}\selectfont{\textbf{Fog}} & \fontsize{6}{7}\selectfont{\textbf{Rain}} & \fontsize{6}{7}\selectfont{\textbf{Sleet}} & \fontsize{6}{7}\selectfont{\textbf{Light snow}} & \fontsize{6}{7}\selectfont{\textbf{Heavy snow}} \\
\hline
\hline
\fontsize{7}{7.2}\selectfont{\multirow{2}{*}{\begin{tabular}[c]{@{}c@{}}RTNH\\ \cite{e-k-radar}\end{tabular}}} & \fontsize{7}{10.1}\selectfont{$\textit{AP}_{\textit{BEV}}$} & \fontsize{7}{10.1}\selectfont{56.7} & \fontsize{7}{10.1}\selectfont{53.8} & \fontsize{7}{10.1}\selectfont{68.3} & \fontsize{7}{10.1}\selectfont{89.6} & \fontsize{7}{10.1}\selectfont{49.3} & \fontsize{7}{10.1}\selectfont{55.6} & \fontsize{7}{10.1}\selectfont{69.4} & \fontsize{7}{10.1}\selectfont{60.3} \\ 
 & \fontsize{7}{10.1}\selectfont{$\textit{AP}_{\textit{3D}}$} & \fontsize{7}{10.1}\selectfont{48.2} & \fontsize{7}{10.1}\selectfont{45.5} & \fontsize{7}{10.1}\selectfont{58.8} & \fontsize{7}{10.1}\selectfont{79.3} & \fontsize{7}{10.1}\selectfont{40.3} & \fontsize{7}{10.1}\selectfont{48.1} & \fontsize{7}{10.1}\selectfont{65.6} & \fontsize{7}{10.1}\selectfont{52.6} \\ 
\hline

\fontsize{7}{10.1}\selectfont{\multirow{2}{*}{\begin{tabular}[c]{@{}l@{}}RTNH-AL\end{tabular}}} & \fontsize{7}{10.1}\selectfont{$\textit{AP}_{\textit{BEV}}$} & \fontsize{7}{10.1}\selectfont{54.7} & \fontsize{7}{10.1}\selectfont{47.1} & \fontsize{7}{10.1}\selectfont{69.4} & \fontsize{7}{10.1}\selectfont{89.2} & \fontsize{7}{10.1}\selectfont{48.5} & \fontsize{7}{10.1}\selectfont{36.5} & \fontsize{7}{10.1}\selectfont{67.5} & \fontsize{7}{10.1}\selectfont{52.0} \\ 
 & \fontsize{7}{10.1}\selectfont{$\textit{AP}_{\textit{3D}}$} & \fontsize{7}{10.1}\selectfont{46.3} & \fontsize{7}{10.1}\selectfont{44.3} & \fontsize{7}{10.1}\selectfont{60.1} & \fontsize{7}{10.1}\selectfont{78.4} & \fontsize{7}{10.1}\selectfont{39.4} & \fontsize{7}{10.1}\selectfont{28.6} & \fontsize{7}{10.1}\selectfont{63.1} & \fontsize{7}{10.1}\selectfont{50.4} \\ 
\hline
\end{tabular}
\vspace{-4mm}
\label{exp_1}
\end{table}

\subsection{Impact of Inaccurate ALs on RTNH-AL Detection Performance}\label{section_exp_2}

In this experiment, we aim to examine the effect of inaccurate ALs in adverse weather conditions on RTNH-AL performance. Further, we explore the generalized training of RTNH-AL by analyzing three RTNH-AL models trained with the subsets of the K-Radar, as mentioned in subsection \ref{3.c}. The $\textit{AP}_{\textit{BEV}}$ and $\textit{AP}_{\textit{3D}}$ of trained RTNH-AL models are presented in Table \ref{exp_2}.

\begin{table}[!h]
\caption{Comparison of various RTNH-AL models trained with weather specific subsets of K-Radar train dataset}
\centering
\begin{tabular}{c*{10}{>{\centering\arraybackslash}p{0.36cm}}}
\hline
\fontsize{6}{7}\selectfont{\textbf{Model}} & \fontsize{6}{7}\selectfont{\textbf{Met- ric}} & \fontsize{6}{7}\selectfont{\textbf{Over- all}} & \fontsize{6}{7}\selectfont{\textbf{Nor- mal}} & \fontsize{6}{7}\selectfont{\textbf{Over- cast}} & \fontsize{6}{7}\selectfont{\textbf{Fog}} & \fontsize{6}{7}\selectfont{\textbf{Rain}} & \fontsize{6}{7}\selectfont{\textbf{Sleet}} & \fontsize{6}{7}\selectfont{\textbf{Light snow}} & \fontsize{6}{7}\selectfont{\textbf{Heavy snow}} \\
\hline
\hline

\fontsize{7}{7.2}\selectfont{\multirow{2}{*}{\begin{tabular}[c]{@{}c@{}}RTNH-AL\\-NO\end{tabular}}} & \fontsize{7}{10.1}\selectfont{$\textit{AP}_{\textit{BEV}}$} & \fontsize{7}{10.1}\selectfont{45.7} & \fontsize{7}{10.1}\selectfont{46.9} & \fontsize{7}{10.1}\selectfont{60.3} & \fontsize{7}{10.1}\selectfont{83.2} & \fontsize{7}{10.1}\selectfont{38.5} & \fontsize{7}{10.1}\selectfont{17.5} & \fontsize{7}{10.1}\selectfont{44.6} & \fontsize{7}{10.1}\selectfont{38.8} \\ 
 & \fontsize{7}{10.1}\selectfont{$\textit{AP}_{\textit{3D}}$} & \fontsize{7}{10.1}\selectfont{36.3} & \fontsize{7}{10.1}\selectfont{43.3} & \fontsize{7}{10.1}\selectfont{56.4} & \fontsize{7}{10.1}\selectfont{61.0} & \fontsize{7}{10.1}\selectfont{29.1} & \fontsize{7}{10.1}\selectfont{6.8} & \fontsize{7}{10.1}\selectfont{36.4} & \fontsize{7}{10.1}\selectfont{34.7} \\ 
\hline

\fontsize{7}{7.2}\selectfont{\multirow{2}{*}{\begin{tabular}[c]{@{}c@{}}RTNH-AL\\-NOFRL\end{tabular}}} & \fontsize{7}{10.1}\selectfont{$\textit{AP}_{\textit{BEV}}$} & \fontsize{7}{10.1}\selectfont{53.7} & \fontsize{7}{10.1}\selectfont{47.2} & \fontsize{7}{10.1}\selectfont{68.8} & \fontsize{7}{10.1}\selectfont{89.0} & \fontsize{7}{10.1}\selectfont{47.7} & \fontsize{7}{10.1}\selectfont{22.4} & \fontsize{7}{10.1}\selectfont{65.4} & \fontsize{7}{10.1}\selectfont{49.3} \\ 
 & \fontsize{7}{10.1}\selectfont{$\textit{AP}_{\textit{3D}}$} & \fontsize{7}{10.1}\selectfont{45.1} & \fontsize{7}{10.1}\selectfont{44.1} & \fontsize{7}{10.1}\selectfont{59.7} & \fontsize{7}{10.1}\selectfont{78.3} & \fontsize{7}{10.1}\selectfont{38.7} & \fontsize{7}{10.1}\selectfont{10.1} & \fontsize{7}{10.1}\selectfont{59.8} & \fontsize{7}{10.1}\selectfont{42.2} \\ 
\hline

\fontsize{7}{7.2}\selectfont{\multirow{2}{*}{\begin{tabular}[c]{@{}c@{}}RTNH-AL\\-ALL\end{tabular}}} & \fontsize{7}{10.1}\selectfont{$\textit{AP}_{\textit{BEV}}$} & \fontsize{7}{10.1}\selectfont{54.7} & \fontsize{7}{10.1}\selectfont{47.1} & \fontsize{7}{10.1}\selectfont{69.4} & \fontsize{7}{10.1}\selectfont{89.2} & \fontsize{7}{10.1}\selectfont{48.5} & \fontsize{7}{10.1}\selectfont{36.5} & \fontsize{7}{10.1}\selectfont{67.5} & \fontsize{7}{10.1}\selectfont{52.0} \\ 
 & \fontsize{7}{10.1}\selectfont{$\textit{AP}_{\textit{3D}}$} & \fontsize{7}{10.1}\selectfont{46.3} & \fontsize{7}{10.1}\selectfont{44.3} & \fontsize{7}{10.1}\selectfont{60.1} & \fontsize{7}{10.1}\selectfont{78.4} & \fontsize{7}{10.1}\selectfont{39.4} & \fontsize{7}{10.1}\selectfont{28.6} & \fontsize{7}{10.1}\selectfont{63.1} & \fontsize{7}{10.1}\selectfont{50.4} \\ 
\hline
\end{tabular}
\vspace{-0mm}
\label{exp_2}
\end{table}

\begin{table}[!h]
\caption{Distribution of Road type in each weather condition subset of K-radar train dataset}
\begin{tabular}{c*{8}{>{\centering\arraybackslash}p{0.6cm}}}
\hline
    \begin{tabular}[c]{@{}c@{}}\fontsize{6}{5.1}\selectfont{\textbf{Weather}}\\ \fontsize{6}{5.1}\selectfont{\textbf{Conditions}}\end{tabular} &
    \begin{tabular}[c]{@{}c@{}}\fontsize{7}{5.1}\selectfont{\textbf{Nor-}}\\ \fontsize{7}{5.1}\selectfont{\textbf{mal}}\end{tabular} &
    \begin{tabular}[c]{@{}c@{}}\fontsize{7}{5.1}\selectfont{\textbf{Over-}}\\ \fontsize{7}{5.1}\selectfont{\textbf{cast}}\end{tabular} &
    \textbf{Fog} &
    \textbf{Rain} &
    \textbf{Sleet} &
    \begin{tabular}[c]{@{}c@{}}\fontsize{7}{5.1}\selectfont{\textbf{Light}}\\ \fontsize{7}{5.1}\selectfont{\textbf{snow}}\end{tabular} &
    \begin{tabular}[c]{@{}c@{}}\fontsize{7}{5.1}\selectfont{\textbf{Heavy}}\\ \fontsize{7}{5.1}\selectfont{\textbf{snow}}\end{tabular} \\
\hline
\hline
\begin{tabular}[c]{@{}c@{}}\fontsize{7}{6.1}\selectfont{Including}\\ \fontsize{7}{6.1}\selectfont{Road}\\ \fontsize{7}{6.1}\selectfont{Types}\end{tabular} &
  \begin{tabular}[c]{@{}c@{}}Urb.,\\ Hwy.,\\ Alley.,\\Univ.\end{tabular} &
  \begin{tabular}[c]{@{}c@{}}Urb.,\\ Hwy.\end{tabular} &
  \begin{tabular}[c]{@{}c@{}}Sub.,\\ Mtn.\end{tabular} &
  \begin{tabular}[c]{@{}c@{}}Urb.,\\ Sub.\end{tabular} &
  \begin{tabular}[c]{@{}c@{}}Sub,\\ Mtn,\\ P-lot.\end{tabular} &
  \begin{tabular}[c]{@{}c@{}}Urb.,\\ Hwy.\end{tabular} &
  \begin{tabular}[c]{@{}c@{}}Urb.,\\ Sub.,\\ Hwy.\end{tabular} \\
\hline
\vspace{-2mm}
\end{tabular}
\fontsize{7}{3}\selectfont{* Urb., Hwy., Alley., Univ., Sub., Mtn., and P-lot refer to urban, highway, alleyway, university, mountain, and parking lot respectively.}
\vspace{-2mm}
\label{kradar-dist}
\end{table}

Contrary to our hypothesis, RTNH-AL-ALL achieves the best detection performance, followed by RTNH-AL-NOFRL and RTNH-AL-NO. In other words, the performance gradually improves as RTNH-AL has experienced more diverse weather conditions, even if multiple inaccurate auto-labeled objects are contained in the subsets of adverse weather conditions. However, it is notable that while RTNH-NO did not experience fog, rain, sleet, light snow, or heavy snow conditions, the network demonstrates a training effect in these conditions except in sleet conditions. Similarly, the RTNH-NOFRL also shows a training effect in untrained heavy snow condition. In addition, it is interesting that RTNH-NOFRL demonstrates a notable improvement in detection performance in heavy snow condition compared to RTNH-NO while they are trained under the same conditions that both RTNH-AL models have not experienced heavy snow conditions at all in the training step.

These results imply that consistent data distribution of  4DRT across various weather conditions enabled the generalized training of RTNH-AL-NO and RTNH-AL-NOFRL in untrained conditions. In addition, we have found another main factor that affects the generalized training detection performance: road distribution. Table \ref{kradar-dist} represents the road distribution of the K-Radar train dataset in each weather condition subset. In the subset of sleet conditions, about half of the labeled objects are captured at parking lots, the road type not present in other weather condition subsets. This unique data distribution may hinder the generalized training of RTNH-AL-NO and RTNH-AL-NOFRL in sleet conditions, thereby demonstrating degraded detection performance as shown in Table \ref{exp_2}. On the other hand, heavy snow conditions involve similar road distributions with normal, overcast, rain, and light snow weather conditions. Therefore, RTNH-AL-NOFRL presents more generalized detection performance than RTNH-AL-NO in the untrained heavy snow conditions, as RTNH-AL-NOFRL has learned more similar road distribution to that of heavy snow conditions subset than RTNH-AL-NO. In summary, the experiment result confirms that both the distribution of weather conditions and road types are crucial for the generalized training of RTNH-AL.

Based on the experiment results, we propose two strategies for the effective expansion of K-Radar using the auto-labeling method. First, it is desirable to capture as much diverse data as possible, even if the generated ALs may contain multiple inaccurate objects in specific weather conditions. Second, in cases where it is challenging to obtain sufficient data under rare adverse weather conditions, an alternative approach could be to focus on collecting a significant volume of data across various road conditions under more typical weather scenarios. This approach is advantageous as we can expect the generalized training of the RTNH-AL across diverse weather conditions.

\subsection{Ablation Studies} \label{section_exp_3}

\subsubsection{Comparison of RTNH-AL models based on detection performance of LONDs for auto-labeling}

Since ALs are generated based on the inference output of trained LODN, the detection performance of the LODN may affect the ALs accuracy and RTNH-AL detection performance. We examined the impact of LODN detection performance on RTNH-AL training using ALs generated by two different LODNs (i.e., SECOND\cite{second} and PVRCNN++\cite{pvrcnn++}). The experiment result is shown in Table \ref{exp_3}.

\begin{table}[!h]
\caption{Comparison of RTNH-AL Models based on the detection performance of LODNs for auto-labeling}
\centering
\begin{tabular}{c*{10}{>{\centering\arraybackslash}p{0.36cm}}}
\hline
\fontsize{6}{7}\selectfont{\textbf{Model}} & \fontsize{6}{7}\selectfont{\textbf{Met- ric}} & \fontsize{6}{7}\selectfont{\textbf{Over- all}} & \fontsize{6}{7}\selectfont{\textbf{Nor- mal}} & \fontsize{6}{7}\selectfont{\textbf{Over- cast}} & \fontsize{6}{7}\selectfont{\textbf{Fog}} & \fontsize{6}{7}\selectfont{\textbf{Rain}} & \fontsize{6}{7}\selectfont{\textbf{Sleet}} & \fontsize{6}{7}\selectfont{\textbf{Light snow}} & \fontsize{6}{7}\selectfont{\textbf{Heavy snow}} \\
\hline
\hline
\fontsize{6}{7.2}\selectfont{\multirow{2}{*}{\begin{tabular}[c]{@{}c@{}}SECOND\\ \cite{second}\end{tabular}}} & \fontsize{7}{10.1}\selectfont{$\textit{AP}_{\textit{BEV}}$} & \fontsize{7}{10.1}\selectfont{70.3} & \fontsize{7}{10.1}\selectfont{70.0} & \fontsize{7}{10.1}\selectfont{80.3} & \fontsize{7}{10.1}\selectfont{90.1} & \fontsize{7}{10.1}\selectfont{61.9} & \fontsize{7}{10.1}\selectfont{60.9} & \fontsize{7}{10.1}\selectfont{79.7} & \fontsize{7}{10.1}\selectfont{52.8} \\ 
 & \fontsize{7}{10.1}\selectfont{$\textit{AP}_{\textit{3D}}$} & \fontsize{7}{10.1}\selectfont{68.4} & \fontsize{7}{10.1}\selectfont{68.4} & \fontsize{7}{10.1}\selectfont{69.1} & \fontsize{7}{10.1}\selectfont{80.7} & \fontsize{7}{10.1}\selectfont{60.0} & \fontsize{7}{10.1}\selectfont{58.7} & \fontsize{7}{10.1}\selectfont{78.0} & \fontsize{7}{10.1}\selectfont{52.2} \\ 
\hline

\fontsize{6}{7.2}\selectfont{\multirow{2}{*}{\begin{tabular}[c]{@{}c@{}}PVRCNN++\\ \cite{pvrcnn++}\end{tabular}}} & \fontsize{7}{10.1}\selectfont{$\textit{AP}_{\textit{BEV}}$} & \fontsize{7}{10.1}\selectfont{74.8} & \fontsize{7}{10.1}\selectfont{74.0} & \fontsize{7}{10.1}\selectfont{87.4} & \fontsize{7}{10.1}\selectfont{88.0} & \fontsize{7}{10.1}\selectfont{70.2} & \fontsize{7}{10.1}\selectfont{64.5} & \fontsize{7}{10.1}\selectfont{84.2} & \fontsize{7}{10.1}\selectfont{55.1} \\ 
 & \fontsize{7}{10.1}\selectfont{$\textit{AP}_{\textit{3D}}$} & \fontsize{7}{10.1}\selectfont{68.5} & \fontsize{7}{10.1}\selectfont{67.6} & \fontsize{7}{10.1}\selectfont{78.4} & \fontsize{7}{10.1}\selectfont{87.1} & \fontsize{7}{10.1}\selectfont{68.1} & \fontsize{7}{10.1}\selectfont{59.4} & \fontsize{7}{10.1}\selectfont{82.5} & \fontsize{7}{10.1}\selectfont{50.1} \\ 
\hline

\hline

\fontsize{6}{7.2}\selectfont{\multirow{2}{*}{\begin{tabular}[c]{@{}l@{}}RTNH-AL-\\ SECOND\end{tabular}}} & \fontsize{7}{10.1}\selectfont{$\textit{AP}_{\textit{BEV}}$} & \fontsize{7}{10.1}\selectfont{54.0} & \fontsize{7}{10.1}\selectfont{51.9} & \fontsize{7}{10.1}\selectfont{67.8} & \fontsize{7}{10.1}\selectfont{88.7} & \fontsize{7}{10.1}\selectfont{41.4} & \fontsize{7}{10.1}\selectfont{34.4} & \fontsize{7}{10.1}\selectfont{68.3} & \fontsize{7}{10.1}\selectfont{51.1} \\ 
 & \fontsize{7}{10.1}\selectfont{$\textit{AP}_{\textit{3D}}$} & \fontsize{7}{10.1}\selectfont{45.9} & \fontsize{7}{10.1}\selectfont{44.2} & \fontsize{7}{10.1}\selectfont{58.2} & \fontsize{7}{10.1}\selectfont{78.9} & \fontsize{7}{10.1}\selectfont{37.8} & \fontsize{7}{10.1}\selectfont{29.3} & \fontsize{7}{10.1}\selectfont{65.6} & \fontsize{7}{10.1}\selectfont{48.8} \\ 
\hline

\fontsize{6}{7.2}\selectfont{\multirow{2}{*}{\begin{tabular}[c]{@{}l@{}}RTNH-AL-\\ PVRCNN++\end{tabular}}} & \fontsize{7}{10.1}\selectfont{$\textit{AP}_{\textit{BEV}}$} & \fontsize{7}{10.1}\selectfont{54.7} & \fontsize{7}{10.1}\selectfont{47.1} & \fontsize{7}{10.1}\selectfont{69.4} & \fontsize{7}{10.1}\selectfont{89.2} & \fontsize{7}{10.1}\selectfont{48.5} & \fontsize{7}{10.1}\selectfont{36.5} & \fontsize{7}{10.1}\selectfont{67.5} & \fontsize{7}{10.1}\selectfont{52.0} \\ 
 & \fontsize{7}{10.1}\selectfont{$\textit{AP}_{\textit{3D}}$} & \fontsize{7}{10.1}\selectfont{46.3} & \fontsize{7}{10.1}\selectfont{44.3} & \fontsize{7}{10.1}\selectfont{60.1} & \fontsize{7}{10.1}\selectfont{78.4} & \fontsize{7}{10.1}\selectfont{39.4} & \fontsize{7}{10.1}\selectfont{28.6} & \fontsize{7}{10.1}\selectfont{63.1} & \fontsize{7}{10.1}\selectfont{50.4} \\ 
\hline
\end{tabular}
\label{exp_3}
\end{table}

In Table \ref{exp_3}, RTNH-AL-PVRCNN++ and RTNH-AL-SECOND refer to the RTNH models trained using ALs generated from PVRCNN++ and SECOND inference outputs, respectively. As shown in the table, PVRCNN++ demonstrates superior performance compared to SECOND. The performance gap between the two LODNs is reflected in the detection performance of RTNH-ALs, with RTNH-AL-PVRCNN++ outperforming RTNH-AL-SECOND. The result demonstrates that the detection performances of LODN, exploited to generate ALs, affect the RTNH-AL detection performances.\\

\subsubsection{Effectiveness of auto-label refinement process on RTNH-AL detection performance} \label{exp_a_2}

In subsection \ref{section_refinement}, we introduced an IoU-based refinement process to refine the inaccurate and missing objects in the ALs. We conducted an ablation study to verify the effect of the refinement, and the result is shown in Table \ref{exp_4}. In the table, RTNH-AL-REF refers to the RTNH model trained with refined ALs, whereas RTNH-AL is trained with unrefined labels. Table \ref{exp_4} verifies the effect of the refinement as the RTNH-AL-REF demonstrates superior performance in all conditions than RTNH-AL.

\begin{table}[!h]
\caption{Comparison of RTNH-AL models based on the implementation of label refinement process}
\centering
\begin{tabular}{c*{10}{>{\centering\arraybackslash}p{0.36cm}}}
\hline
\fontsize{6}{7}\selectfont{\textbf{Model}} & \fontsize{6}{7}\selectfont{\textbf{Met- ric}} & \fontsize{6}{7}\selectfont{\textbf{Over- all}} & \fontsize{6}{7}\selectfont{\textbf{Nor- mal}} & \fontsize{6}{7}\selectfont{\textbf{Over- cast}} & \fontsize{6}{7}\selectfont{\textbf{Fog}} & \fontsize{6}{7}\selectfont{\textbf{Rain}} & \fontsize{6}{7}\selectfont{\textbf{Sleet}} & \fontsize{6}{7}\selectfont{\textbf{Light snow}} & \fontsize{6}{7}\selectfont{\textbf{Heavy snow}} \\
\hline
\hline

\fontsize{6}{5.2}\selectfont{\multirow{2}{*}{\begin{tabular}[c]{@{}c@{}}RTNH-AL\end{tabular}}} & \fontsize{7}{10.1}\selectfont{$\textit{AP}_{\textit{BEV}}$} & \fontsize{7}{10.1}\selectfont{54.7} & \fontsize{7}{10.1}\selectfont{47.1} & \fontsize{7}{10.1}\selectfont{69.4} & \fontsize{7}{10.1}\selectfont{89.2} & \fontsize{7}{10.1}\selectfont{48.5} & \fontsize{7}{10.1}\selectfont{36.5} & \fontsize{7}{10.1}\selectfont{67.5} & \fontsize{7}{10.1}\selectfont{52.0} \\ 
 & \fontsize{7}{10.1}\selectfont{$\textit{AP}_{\textit{3D}}$} & \fontsize{7}{10.1}\selectfont{46.3} & \fontsize{7}{10.1}\selectfont{44.3} & \fontsize{7}{10.1}\selectfont{60.1} & \fontsize{7}{10.1}\selectfont{78.4} & \fontsize{7}{10.1}\selectfont{39.4} & \fontsize{7}{10.1}\selectfont{28.6} & \fontsize{7}{10.1}\selectfont{63.1} & \fontsize{7}{10.1}\selectfont{50.4} \\ 
\hline

\fontsize{6}{5.2}\selectfont{\multirow{2}{*}{\begin{tabular}[c]{@{}c@{}}RTNH-AL\\-REF\end{tabular}}} & \fontsize{7}{10.1}\selectfont{$\textit{AP}_{\textit{BEV}}$} & \fontsize{7}{10.1}\selectfont{55.6} & \fontsize{7}{10.1}\selectfont{47.9} & \fontsize{7}{10.1}\selectfont{69.5} & \fontsize{7}{10.1}\selectfont{89.9} & \fontsize{7}{10.1}\selectfont{48.9} & \fontsize{7}{10.1}\selectfont{34.6} & \fontsize{7}{10.1}\selectfont{76.6} & \fontsize{7}{10.1}\selectfont{52.5} \\ 
 & \fontsize{7}{10.1}\selectfont{$\textit{AP}_{\textit{3D}}$} & \fontsize{7}{10.1}\selectfont{47.6} & \fontsize{7}{10.1}\selectfont{45.6} & \fontsize{7}{10.1}\selectfont{66.4} & \fontsize{7}{10.1}\selectfont{79.3} & \fontsize{7}{10.1}\selectfont{40.3} & \fontsize{7}{10.1}\selectfont{30.5} & \fontsize{7}{10.1}\selectfont{66.3} & \fontsize{7}{10.1}\selectfont{51.6} \\ 
\hline
\end{tabular}
\vspace{-3mm}
\label{exp_4}
\end{table}

%%%%%%%%%%%%%%%%%%%%%%%%%%%%%%%%%%%%%%%%%%%%%%%%%%%%%%%%%%%%%%%%%%%%%%%%%%%%%%%%
%%%%%%%%%%%%%%%%%%%%%%%%%%%%%%%%%%%%%%%%%%%%%%%%%%%%%%%%%%%%%%%%%%%%%%%%%%%%%%%%

\section{Conclusion}
In this paper, we have proposed an auto-labeling method for 4DRT using a LODN trained with calibrated LPC. The proposed method generated ground truth labels for RTNH training without additional human aid or expense. The experiment results have demonstrated the effectiveness of the proposed method as the RTNH trained with ALs achieved similar detection performance to the RTNH trained with HLs. In addition, we have provided the strategies for efficient expansion of the K-Radar dataset based on the results of the experiment using generated ALs.

\vspace{2mm}
\section*{ACKNOWLEDGMENT}
This work was supported by the National Research Foundation of Korea(NRF) grant funded by the Korea government(MSIT) (No. 2021R1A2C3008370).
\vspace{2mm}
\bibliographystyle{IEEEtran}
\bibliography{reference}

\end{document}